\documentstyle[12pt]{article}

\font\tenrm=cmr10

\begin{document}

\newcommand{\beq}{\begin{equation}}
\newcommand{\eeq}{\end{equation}}
\newcommand{\beqa}{\begin{eqnarray}}
\newcommand{\eeqa}{\end{eqnarray}}
\newcommand{\beqar}{\begin{eqnarray*}}
\newcommand{\eeqar}{\end{eqnarray*}}
\newcommand{\tr}{{\rm tr}}
\newcommand{\be}{\beta}
\newcommand{\al}{\alpha}
\newcommand{\ie}{{\it i.e.,}\ }
\newcommand{\eg}{{\it e.g.,}\ }
\newcommand{\ch}{{\cal H}}
\newcommand{\ssc}{\scriptscriptstyle}
\def\d{\delta}
\def\eps{\epsilon}
\def\heps{\hat{\epsilon}}
\def\beps{\bar{\epsilon}}
\def\rL{\widetilde{L}}
\def\tc{\tilde{\chi}^a}
\def\tchi{\tilde{\chi}}
\def\ddx{d^4\!x}
\def\dtx{d^{2}\!x}
\def\S{{\cal S}}
\def\ls{{\ell_s}}
\def\g{{g_s}}

\renewenvironment{thebibliography}[1]
  { \begin{list}{\arabic{enumi}.}
    {\usecounter{enumi} \setlength{\parsep}{0pt}
     \setlength{\itemsep}{3pt} \settowidth{\labelwidth}{#1.}
     \sloppy
    }}{\end{list}}

\parindent=1.5pc

 ~
\vskip -0.5truein
\rightline{\small \hfill gr--qc/0107034} 
\vskip 0.4truein


\begin{center}
{{\large \bf BLACK HOLES AND STRING THEORY}\\
\vglue 1.0cm
{ROBERT C. MYERS}\\[1em]
\baselineskip=14pt
{\it Department of Physics, McGill University}\\
\baselineskip=14pt
{\it Montr\'eal, Qu\'ebec, Canada H3A 2T8}\\


\vglue 0.8cm
{\tenrm ABSTRACT}}
\end{center}
{\rightskip=3pc
 \leftskip=3pc
 \tenrm\baselineskip=12pt
 \noindent
This is a short summary of my lectures given at the Fourth Mexican School
on Gravitation and Mathematical Physics. These lectures gave a 
brief introduction
to black holes in string theory, in which I primarily focussed on describing
some of the recent calculations of black hole entropy using the
statistical mechanics of D-brane states. The following overview will
also provide the interested students with an introduction to the
relevant literature.

\vglue 0.8cm}

%
%
{\bf\noindent 1. Prologue}
\vglue 0.4cm
\baselineskip=14pt
String theory is a very broad and extremely rich area
of study in theoretical physics pursued by particle physicists,
mathematicians, and relativists as well. Within this community of
string theorists, there has long been a fascination with black holes, and
studies of the latter have taken many different points of view, including:
\begin{enumerate}
\setlength{\itemsep}{-1mm}
\setlength{\parsep}{0mm}
\item Black holes with string theory corrections \cite{cmp}
\item Black holes with quantum hair \cite{cpw}
\item Two-dimensional black holes as WZW models \cite{wv}
\item Black holes in solvable models of two-dimensional gravity \cite{hsg}
\item Black holes and entanglement entropy \cite{suss}
\item Black holes as low energy supergravity solutions \cite{youm}
\item Black holes as exact sigma model backgrounds \cite{test}
\item Black holes as strings \cite{callan}
\item Black holes as D-branes \cite{juan,peet,dama}
\item Black holes in Matrix theory \cite{banks}
\item Black holes in the AdS/CFT correspondence \cite{magoo}
\item Black holes as superconformal quantum mechanics \cite{andy}
\item Black holes in brane world scenarios \cite{world}
\item Black holes and enhan\c con physics \cite{enhan}
\end{enumerate}
The references cited above are by no means complete. Consulting
Paul Ginsparg's e-print archive \cite{arch}, one finds that in the past ten
years, the high energy theory (hep-th)
section has accumulated in excess of 1600 papers about black holes.
Above, I have only listed a few reviews
or salient articles for each topic to give the reader a bridgehead into the
relevant literature. The interested students are encouraged to
explore the associated references and citations of these papers
with the Spires HEP database \cite{spires}. 

Clearly, I could not hope to tell the full story of black holes
and string theory in two hour-long lectures. Instead I only attempted
to introduce the students to the ninth item on the list above. That is,
I described some of the recent calculations of black hole entropy using
techniques involving D-branes. In particular, I focussed on the original
calculations of Strominger and Vafa \cite{vafa}. These were the
first calculations of any sort which successfully determined the
Bekenstein-Hawking entropy with a statistical mechanical model
in terms of some underlying microphysical states. 
There are already several extensive reviews of the D-brane description of
black hole microphysics. In particular,
I would recommend those by Peet \cite{peet} and by Das and Mathur \cite{dama}.
I would also highly recommend Juan Maldacena's Ph.D. thesis \cite{juan}
as a well-written and pedagogical introduction to this topic. With regards
to further background references, Clifford Johnson's review \cite{primer} of
D-brane physics is very good. For a general introduction to modern string
theory, the standard reference is now Polchinski's text \cite{joep}.
Interested students may also wish to look at a similar
but longer series of lectures on black holes in string theory,
which I presented in Jerusalem in the previous year \cite{jeru}.

\vglue 0.6cm
{\bf \noindent 2. Summary}
\vglue 0.4cm

In the early seventies, Bekenstein \cite{bek} made the bold conjecture
that black holes carry an intrinsic entropy given by the surface area of the
horizon measured in Planck units multiplied by
a dimensionless number of order one. In part,
this conjecture was motivated by Hawking's area
theorem \cite{areathm} which had shown that, like entropy, the horizon area
of a black hole can never decrease in general relativity.

The next crucial insight came from Hawking while investigating quantum fields
in a black hole spacetime \cite{radi}. He found that
external observers detect the emission of thermal radiation from
a black hole with a temperature proportional to its surface
gravity\footnote{The surface gravity may be thought of as the redshifted
acceleration of a fiducial observer moving just outside the horizon
\cite{wtext}.},
$\kappa$:
\beq
k_{\scriptscriptstyle B}T={\hbar\kappa\over2\pi c}\ \ .
\label{temp}
\eeq
For a Schwarzschild black hole, $\kappa=c^4/(4GM)$
and so one finds that Hawking's result typically
corresponds to an incredibly small
temperature: $ T \sim 10^{-7} K$ for a solar mass black hole.

Previously, extensive studies of solutions of Einstein's equations
had culminated in the formulation of four laws of
black hole mechanics\cite{barcar}.
Hawking's discovery of a black hole temperature
was the key to realizing that these previous results were the laws
of thermodynamics applied to black holes.
For example, there is a correspondence between the first law
in each of these frameworks:
\begin{equation}
\frac{c^2\kappa}{8\pi G}\delta A=c^2\delta M\quad
\longleftrightarrow\quad T\delta S=\delta U\ .
\label{first}
\end{equation}
Here, $Mc^2$ is naturally identified with
the black hole's internal energy, $U$. Hence given Hawking's
relation (\ref{temp}) between  the
surface gravity and the  temperature, the correspondence
between these two relations is completed by identifying \cite{radi}
\beq
S={k_{\scriptscriptstyle B}c^3\over\hbar\, G}\,{A\over 4}\ ,
\label{entropy1}
\end{equation}
which gives a precise relation confirming Bekenstein's earlier
conjecture.

However, these revelations about the thermal nature of black
holes lead to two related puzzles. The above discussion describes
black hole entropy within the framework of thermodynamics,
where it is associated with the energy in
a system which is unavailable to do work, \eg in eq.~(\ref{first}),
$T\,\delta S$ indicates the heat loss in some process.
For ordinary thermal systems, statistical mechanics provides
a complementary interpretation of entropy by taking into account
the microscopic degrees of freedom of the system.
In this context, entropy has quite a different significance. It
is a measure of the lack of detailed
information about the microphysical state of
a system.  However, in the case of black holes, it remained a
longstanding problem to find a statistical mechanical derivation
of the entropy.

An even more dramatic puzzle is the black hole information loss paradox.
Classical general relativity says that whatever falls into a black hole
cannot afterwards be observed from the outside. In principle though,
we could discover what fell in by entering the black hole ourselves.
However if quantum  processes
cause the black hole to radiate away its energy thermally so that
eventually the
black hole disappears, then the information about what has fallen
in is completely lost. In fact,
such a loss of information violates unitary time evolution, one of
the basic tenets of quantum mechanics,
the theory which lead to the black hole evaporation
in the first place. This paradox has profound implications as it was
originally suggested to indicate that quantum mechanics
and general relativity simply can not be combined in a
consistent manner. It was long felt that a resolution of either
of these puzzles would yield  some insight into the nature of quantum gravity.
This is the essential source of the fascination which string
theorists and particle physicists have for black holes.

Recently, progress into these questions has been made with new
insights from string theory. This progress is a spin-off from
the research into string dualities \cite{dual} and the realization of the
important role of extended objects beyond just strings \cite{joed}.
In particular, a class of extended objects known as 
Dirichlet branes or D-branes \cite{primer}
have proven very valuable from a calculational standpoint.
These objects have a simple description in the framework
of perturbative or weakly-interacting strings, and yet they exhibit
rich dynamics, including a wide variety of complicated bound states.

In the low energy or long wavelength limit, string theory is accurately
described by Einstein gravity coupled to various kinds of matter. 
In my lectures, I focussed on what is known as the Type IIb superstring
theory. In this case, one has a ten-dimensional supergravity theory
where the matter fields include two scalars (the dilaton
and the axion), two two-form potentials, a four-form potential
and various fermions. As we saw in Don Marolf's lectures \cite{donl},
various kinds of extended objects can carry charges under the form fields.
The fundamental strings of the theory act as the electric
sources for one of the two-form potentials, known as the NS (Neveu-Schwarz)
two-form. The other potential, the
RR (Ramond-Ramond) two-form, has electric sources known as D1-branes,
and magnetic sources known as D5-branes (\ie these are
D-branes extended in one and five spatial dimensions, respectively).
From the full quantum string theory, we know there is an analog of
Dirac charge quantization for ordinary electric charges and magnetic monopoles
in four dimensions \cite{mag}, which requires that the RR two-form charges
come in discrete units \cite{joed}. Hence if a system carries a certain RR
charge, one
can use the charge to count the total number of constituent D-branes that
must have been used in assembling the system.

In the lectures, I focussed on a particular family of black hole
solutions in the Type IIb supergravity theory.
From a ten-dimensional point of view, these solutions describe black
five-branes carrying three distinct types of charge, including
both electric and magnetic charge with respect to the RR two-form.
Hence we can say that the black brane was formed by bringing together
some number of D1-branes and D5-branes, $N_1$ and $N_5$. 
Furthermore, the details of the solution allow us to infer
that all of the D5-branes were arranged in parallel  on a common
five-dimensional hypersurface and that,
similarly, the D1-branes were parallel on a common line in this surface.
In order that the resulting black brane has a finite mass, charge and horizon
area, we imagine that
the directions in the above hypersurface are wrapped on circles to
form a five-dimensional torus. The third charge is a momentum along the
circle common to the D1-branes and D5-branes. A standard result of
KK (Kaluza-Klein) theory is that such an internal momentum
must also be quantized \cite{KK}, and so we use $N_P$ to denote the number of
momentum quanta carried by the solution. From the point of view of the
effective five-dimensional theory, the Penrose diagram for these solutions
is similar to that of the Reissner-Nordstrom solution in four
dimensions \cite{wtext}. The solution presented
is distinguished by the fact that there is a supersymmetric limit
in which the horizon area remains finite and the black brane becomes
extremal (\ie the surface gravity and hence the Hawking temperature
vanish). Identifying black hole solutions with these properties
is nontrivial as can be seen by the fact that if any
of the three charges is set to zero, the horizon is replaced by
a null singularity in the supersymmetric limit.
Evaluating the black hole entropy according to the Bekenstein-Hawking
formula (\ref{entropy1}) yields
\beq
S=2\pi\sqrt{N_1 N_5 N_P}\ .
\label{entropya}
\eeq
Note that the right hand side is a pure number that does not depend,
\eg on the details of the compactification to five dimensions.

This result (\ref{entropya}) for the classical supergravity solution relies
on two inputs from the underlying Type IIb string theory. The first was
the charge quantization conditions alluded to above,
and the second was a formula for
Newton's constant in ten dimensions: $16\pi G=(2\pi)^7\g^2\ls^8$.
Here Newton's constant is expressed in terms of two parameters
arising in the perturbative string theory: $\g$, the string coupling
constant, a dimensionless parameter which describes the strength with
which the fundamental (closed) strings interact,
and $\ls$, the string scale which can be regarded as
the typical size of a fundamental string.

As mentioned previously, the D-branes can also be analysed using the
techniques of perturbative string theory. From this point of view,
one is considering a particular bound state of $N_1$ D1-branes
and $N_5$ D5-branes. The $N_P$ units of momentum are carried by
fundamental strings connecting the D1-branes and D5-branes. It should
be evident that there are of the order of $N_1N_5$ different species of
strings that can serve in this role, and further that
this momentum may be partitioned amongst the various string excitations
in many different ways, \ie an individual string may carry anywhere between
1 and $N_P$ units of momentum. Therefore the supersymmetric ground state
of this bound state has a large degeneracy, $\cal D$. Given this
degeneracy, one can assign a statistical mechanical entropy to the system
according to $S=\log{\cal D}$. A precise evaluation of the degeneracy
then exactly reproduces the entropy given in eq.~(\ref{entropya}).
Hence this calculation yields a striking agreement between the
Hawking-Bekenstein entropy and the statistical entropy of the
D-brane microstates.

Now at this point, the attentive student must have been asking:
what does a calculation
of the degeneracy of a D-brane bound state have to do with a calculation
of the Hawking-Bekenstein entropy of a black hole solution?
The relation between these calculations is that the D-brane bound state
in perturbative string theory and the black hole
in supergravity are actually complementary descriptions of the
same system, valid in different regimes of the coupling. Despite their
simple description in perturbative string theory, D-branes are
nonperturbative objects. This can be seen from the tension
(energy density) for a single D-brane which is inversely
proportional to the string coupling,  $T\sim 1/\g$. However,
the gravitational footprint may still be small as this involves
coupling the D-brane to gravity with Newton's constant. With the
result quoted above, one finds, for example, for a collection of
D5-branes: $r_g^{2} =G N_5 T\sim N_5\g\ls^{2}$.
This radius $r_g$ can be regarded as the length scale
over which the curvatures and other fields are strong.
Hence in a regime where $N_5\g\ll 1$, $r_g$ is less than the
string scale $\ls$ and so is a distance that we can't expect to resolve
effectively in the perturbative string theory. In this regime, the
D-branes are effectively described by a perturbative picture where
the D-branes are represented by a source in flat empty space which
couples weakly to the fundamental strings. On the other hand,
in the regime $N_5\g\gg 1$, $r_g$ is much larger than the
string scale $\ls$ and a nonlinear supergravity solution providing
a background for the propagation of the strings is a better description
of the physics. Note that this ``strong coupling" regime may still
have $\g\ll 1$. Then the fundamental strings couple weakly to each other
but interact strongly with the collection of D-branes.

Therefore the perturbative string picture and the black hole solution
provide complementary pictures of the same D1-D5 bound state. One
point of view is that when $\g$ (and hence Newton's constant) is increased
above the weak coupling regime, the
system undergoes gravitational collapse forming a black hole.
The final step in the argument is that the system under consideration
is supersymmetric \cite{susy}. Supersymmetry plays
an essential role here in that it guarantees that the number of ground
states is independent of the strength of the string coupling, \ie
the ground state degeneracy is a kinematic quantity rather than a
dynamical quantity. Therefore one can expect that the
calculations in both regimes will yield the same result.

The original calculations of the five-dimensional black hole \cite{vafa}
were quickly extended to spinning black holes \cite{jake}, four-dimensional
black holes \cite{four} and also near-extremal black holes \cite{near}.
In the latter case, where the temperature is slightly greater than zero,
one can develop a D-brane model for the
process of Hawking evaporation \cite{evap}. This microscopic
model \cite{evap2} even captures the grey body factors which modify the
thermal spectrum for the particles radiated to asymptotic infinity \cite{grey}.
For the near-extremal calculations, one has lost supersymmetry and so one must
modify the arguments which relate the results in the weak and strong
coupling regimes \cite{juan2}. Further, the robustness of these D-brane
models is related to the AdS/CFT correspondence \cite{magoo},
which comes into play in describing the near horizon region of the
five-dimensional black holes.

These results represented a major breakthrough in our
understanding of black hole microphysics, as a statistical mechanical
interpretation of black hole entropy had eluded theoretical
physicists for over 20 years following the discovery of Hawking radiation.
While the paradox of information loss in black hole evaporation
remains unresolved,  D-branes seem to provide
a robust model of at least certain evaporating black holes and
so the tools to resolve this perplexing paradox seem to be at hand.
In any event, the present remarkable calculations already
provide further sanction for string theory as the theory to
reconcile quantum mechanics and general relativity.

\vglue 0.6cm
{\bf \noindent Acknowledgements \hfil}
\vglue 0.4cm
We would like to congratulate the organizers of the Fourth Mexican School
on Gravitation and Mathematical Physics for arranging a very successful
gathering. I would also like to thank them for giving me the opportunity
to lecture at their school, as well as enjoying the pleasant surroundings of
Huatulco. This work was supported in part by NSERC of Canada and Fonds FCAR
du Qu\'ebec. I would to thank the Institute for Theoretical Physics at
UCSB for hospitality during the latter stages of writing these notes.
Research at the ITP was supported in part by the U.S.  National
Science Foundation under Grant No.  PHY99--07949.
Finally I would like to thank Neil Constable, Frederic Leblond
and David Winters for carefully reading an earlier draft of these notes.

\vglue 0.6cm
{\bf\noindent References \hfil}
\vglue 0.4cm

\end{document}